\documentclass[10pt,leqno]{article}
\usepackage{graphicx}
\usepackage{indentfirst,csquotes}

\usepackage{amssymb,amsthm,amsmath}
\usepackage{xcolor,paralist,hyperref,titlesec,fancyhdr,etoolbox}

\hypersetup{ colorlinks=true, linkcolor=black, filecolor=black, urlcolor=black }

\usepackage{lipsum}
\usepackage{epsfig}
\usepackage{algorithm}
\usepackage{algpseudocode}
\usepackage{caption}
\usepackage{subcaption}

\begin{document}
\title{IOD\emph{eep}: an IOD for the introduction of deep learning in the DICOM standard} 
\author{Salvatore Contino$^{1, \ddagger}$, Luca Cruciata$^{1, \ddagger}$, Orazio Gambino$^{1, *}$, Roberto Pirrone$^{1}$}

\maketitle
\let\thefootnote\relax
\footnotetext{$^{*}$corrisponding author orazio.gambino@unipa.it (Orazio Gambino)} 
\footnotetext{$^{\ddagger}$These authors contributes equally.} 
\footnotetext{$^{1}$Department of Engineering, University of Palermo, 90128 Palermo, Italy} 

\begin{abstract}
\textbf{Background and Objective:} In recent years, Artificial Intelligence (AI) and in particular Deep Neural Networks (DNN) became a relevant research topic in biomedical image segmentation due to the availability of more and more data sets along with the establishment of well known competitions.
Despite the popularity of DNN based segmentation on the research side, these techniques are almost unused in the daily clinical practice even if they could support effectively the physician during the diagnostic process. Apart from the issues related to the explainability of the predictions of a neural model, such systems are not integrated in the diagnostic workflow, and a standardization of their use is needed to achieve this goal.
\textbf{Methods:} This paper presents \textit{IODeep} a new DICOM Information Object Definition (IOD) aimed at storing both the weights and the architecture of a DNN already trained on a particular image dataset that is labeled as regards the acquisition modality, the anatomical region, and the disease under investigation. \textbf{Results:} The IOD architecture is presented along with a DNN selection algorithm from the PACS server based on the labels outlined above, and a simple PACS viewer purposely designed for demonstrating the effectiveness of the DICOM integration, while no modifications are required on the PACS server side. Also a service based architecture in support of the entire workflow has been implemented. \textbf{Conclusion:} IODeep ensures full integration of a trained AI model in a DICOM infrastructure, and it is also enables a scenario where a trained model can be either fine-tuned with hospital data or trained in a federated learning scheme shared by different hospitals. In this way AI models can be tailored to the real data produced by a Radiology ward thus improving the physician decision making process. 
\\Source code is freely available at https://github.com/CHILab1/IODeep.git
\end{abstract} 

\section{Introduction}\label{intro} 
As reported in the most recent scientific literature, state-of-the-art techniques for segmenting biomedical images make use of AI, and in particular Deep Learning (DL) models that  provide an average accuracy higher than \(90\%\) on most tasks.
Nevertheless, DL models are almost no present in the daily diagnostic practice, due to three main problems stated below:
\begin{itemize}
    \item Statement 1: DL models are not integrated in the medical images diagnostic platforms.
    \item Statement 2: DL models can not be trained on huge data sets coming from radiologic wards due to issues related to both the patient's data privacy and the security issues implied in building a computing infrastructure distributed across several hospitals.
    \item Statement 3: Physicians do not trust the predictions by a neural model as they do not understand the ``clinical way" used by the model to segment regions that are relevant for the diagnosis.
\end{itemize}

Despite the problems stated above, using DL models in biomedical imaging platforms would bring multiple benefits:
\begin{itemize}
  \item Improve efficiency: DL models can automate routine tasks such as image segmentation and analysis, freeing up radiologists' time and improving workflow efficiency.
  \item Cognitive and visual fatigue decreasing: The radiologists' cognitive and visual fatigue entail the wrong diagnosis, whereas an AI algorithm is not affected by this. As a consequence, AI is a valid support during the image analysis by the radiologist \cite{WAITE2017191}. 
  \item Increase accuracy: DL models can help to identify very subtle patterns in medical images that may be missed by the human eye, leading to more accurate and reliable diagnoses \cite{Savage2020}.
  \item Enhance the quality of care: AI can provide additional information to support clinical decision-making, helping healthcare providers to make more informed and effective treatment decisions \cite{Rundo2020}.
  \item Facilitate research: AI based techniques can help to automate the analysis of large data sets, enabling researchers to identify new patterns and insights that may not be apparent through traditional methods.
  \item Mass screening: the ability to analyze large data sets provided by AI algorithms can be used to perform mass screening on the occasion of campaigns to raise awareness of a particular illness \cite{UKHAROV2023101223}.
  \item Improve patient outcomes: by improving the speed and accuracy of diagnoses, DL models can help to improve patient outcomes and reduce the need for follow-up testing.
\end{itemize}

In view of the previous considerations, this work provides a complete solution to the problem formulated in Statement 1, and it also represents an enabling technology for answering both Statement 2 and Statement 3.  
The system proposed in our work leverages the well known DICOM (Digital Imaging and Communications in Medicine)\footnote{\url{https://www.dicomstandard.org/}} standard to achieve the solution, by integrating in the normal DICOM communication between a PACS server and a viewer, the storage and retrieval of a DNN trained purposely on a data set that is related to a particular acquisition modality, anatomical region, and disease. We achieve our stated goal through the creation of a proper Information Object Definition (IOD) that we called \emph{IODeep}. IODeep contains both the architecture and the weights of a neural network purposely trained on a medical image data set; in our approach, several DNNs are stored in the PACS server, each in a separate IODeep instance, and a novel algorithm is proposed to select the best matching DNN for the images under investigation, based on the information mentioned above that IODeep stores in some proper DICOM tags. Network selection is transparent for the physician who only asks the system to start AI based Regions of Interest (ROI) proposal. The selected network is used to perform client-side prediction of the ROI on the slice that the physician is currently looking at. On the PACS viewer side, ROIs are visually proposed to the physician who can explicitly validate them as relevant for the diagnosis. The PACS client embeds the set of validated ROIs in an RT Structure Set IOD, which is stored on the server. The entire workflow is conducted in accordance with the DICOM protocol, preserving the security required by the standard, and no modification is required on the PACS server side. The main contributions of the paper are the following:

\begin{itemize}
    \item The complete information architecture of IODeep along with the DICOM compliant workflow for both ROI prediction and storage on the PACS server.
    \item A novel algorithm based on the IODeep DICOM tags to select the correct DNN for the data under investigation in a transparent way for the physician.
    \item A purposely designed back-end  independent JSON format to describe DNN architectures to be used as a possible DNN description inside IODeep.
    \item A purposely designed lightweight PACS viewer that implements all the client side communication involved in using IODeep, while no modification is required on the server side.
    \item A service based implementation of the entire ROI prediction scenario where both network selection and prediction run as external REST services that communicate with the PACS infrastructure
\end{itemize}

The rest of the paper is arranged as follows. Section \ref{sec:back} reports some theoretical background about DL models for medical image analysis. The motivations of our work are outlined in Section \ref{sec:motiv}, while the IODeep architecture is described in Section \ref{sec:arch} along with the detailed ROI prediction workflow, and a comparison with the current abstract model devised by the DICOM community for AI integration. Final remarks and conclusions are reported in Section \ref{sec:conc}.

\section{Theoretical Background}\label{back}
A relevant amount of recent papers in the scientific literature on biomedical image processing make large use of deep-learning approaches. Such types of algorithms are applied without any regard for image modality, pathology and body part with the aim to either segment or simply classify both 2D and 3D data. DNNs are used mainly in three kinds of task: classification, semantic segmentation, and region proposal.

Classification tasks aim at placing a (multi-)class label to the whole image, as in the case of histopathology where one is interested in assessing the presence of cancer tissues. Conversely, semantic segmentation is the task of labeling each pixel in the image as belonging to a certain class. In general, labels discriminate different tissues: a normal organ, a pathological one, a tumor lesion, and so on. Zhou et al. \cite{zhou2018unet++} propose a nested U-Net architecture for medical image segmentation. It soon achieved state-of-the-art results on several data sets including lung and liver CT scans. Schlemper et al. \cite{schlemper2019attention} introduce an attention mechanism to the U-Net architecture for pancreas segmentation in CT scans. The attention mechanism helps the model to focus on the relevant regions of the image, leading to improved segmentation accuracy. The approach proposed by Fu et al. \cite{fu2019dual} proposes a dual attention mechanism that has been adapted for medical image segmentation tasks with good results. The mechanism consists of a spatial attention module and a channel attention module, which capture spatial and channel-wise dependencies respectively. Dou et al. \cite{dou2020anisotropic} propose a 3D anisotropic hybrid network for medical image segmentation, which combines 2D and 3D convolutional layers to better handle anisotropic medical images. Milletari et al. \cite{milletari2021vnet} introduce a family of V-Net architectures for volumetric medical image segmentation, which achieve state-of-the-art results on several data sets including liver, lung, and brain CT scans. The V-Net architectures are based on 3D convolutional layers and incorporate skip connections for improved segmentation accuracy. Christ et al. \cite{christ2017automatic} propose a cascaded fully convolutional neural network (FCN) approach for automatic segmentation of liver and tumor from CT and MRI volumes. The proposed method consists of two stages, where the first stage segments the liver, and the second stage segments the tumor within the liver. The approach achieved state-of-the-art results on two liver tumor segmentation datasets. A review on the state of the art in deep learning-based methods for brain MRI segmentation is presented in \cite{Akkus2017}. The authors provide an overview of different types of deep learning models used for brain MRI segmentation, including fully convolutional networks, U-Nets, and deep residual networks. The paper also discusses current challenges and future directions for brain MRI segmentation. Qu et al. \cite{Qu2021} present a deep learning-based method to segment pelvic bone tumors in MRI that is based on a a multi-view fusion network to extract pseudo-3D information from two scans in different directions. Wang et al. \cite{Wang2019} propose a novel deep neural architecture for segmenting multiple organs in abdominal CT scans, that is called Organ-Attention Network (OAN), and it is specifically designed to both classify and segment each individual organ with high accuracy. OAN uses a combination of convolutional and attention-based modules to selectively focus on each organ of interest while ignoring irrelevant regions in the image. Additionally, the paper proposes a statistical fusion strategy to combine the individual organ segmentations generated by OAN into a complete multi-organ segmentation. Yan and colleagues \cite{Yan2021} propose a deep learning approach for the automatic segmentation of pancreas from computed tomography (CT) scans. The proposed method uses a multi-scale U-Net architecture, tested on NIH pancreas segmentation data set  \cite{roth2015deeporgan}, which includes an attention mechanism to highlight important regions in the input image. A very recent attention based CNN is a module based on Cortical Spiking models presented by Zhou et al. \cite{Zhou_Huang_Ding_Zhang_2023}, that reduces the number of trainable parameters compared to well-known state-of-the-art architectures, while improving performance in terms of accuracy. 
In very recent times, the advent of Transformers networks, and self-attention mechanisms originated in the Natural Language Processing (NLP) community \cite{46201} has pushed also the research in Computer Vision towards the use of networks integrating this kind of technology. Vision transformers (ViTs) have become attractive due to their ability to encode long-range dependencies and to learn highly effective feature representations. Right now, ViTs are the new state of the art, surpassing CNN-based architectures in both classification and semantic segmentation tasks \cite{Dosovitskiy_2021}, adding more and more trainable parameters but decreasing the possibility of gaining information from eXplainable AI (XAI) techniques.

CNN-based networks are often used as a backbone for architectures trained to solve region proposal tasks. In fact, in contrast to semantic segmentation that needs pixel precision, region proposal architectures return predictions about ROIs, i.e. bounding boxes around the truly interesting pattern. 
Punn and Agarwal \cite{Punn_Agarwal_2022} in their 2022 survey analysed the state-of-the-art of all U-Net variants used in region proposal tasks; the survey's emphasis is on the efficiency of the various variants in tricky tasks such as the diagnosis of severe acute respiratory syndrome coronavirus 2 (SARS-CoV-2). 
Zhang et al. in 2023 \cite{Zhang_Zhang_Jin_Xu_Xu_2023} propose the MDU-Net, an architecture consisting of a multi-scale dense connections (MDC) encoder and a U-shaped decoder for biomedical image segmentation. The multi-scale dense connections, allowing shorter connections between layers close to the input and output, enable a much deeper U-net that reduces potential overfitting and further improves segmentation performance. 
Finally, Song et al. \cite{Song_Wang_Zeng_Guo_Li_2023} in their work propose the Outlined Attention U-Network, (OAU-Net) based on a bypass branching strategy to solve biomedical image segmentation tasks, capable of detecting surface and deep features. The particularity of this architecture lies in the encoders that are based on residual convolution.

One of the main disadvantages of region proposal architectures is the high computational cost deriving from the increasing number of trainable parameters.  In very recent years, Weakly Supervised Object Location (WSOL) and Weakly Supervised Semantic Segmentation (WSSS) have gained relevance. Weakly supervised techniques do not need explicit explicit pixel-wise labeling to be trained. WSOL~\cite{DBLP:journals/corr/abs-2104-07918} is the task of learning how to place a bounding box or a loose ROI around a relevant object in the image, according to the label provided at image level. Many kinds of WSSS~\cite{shen2023survey} are reported in the literature. The most common task in this respect is learning explicit (pixel-wise) segmentation of the object starting from either an image-wise or a box-wise label. Some recent WSSS approaches leverage the Class Activation Maps (CAM)~\cite{7780688}. CAM is a well known XAI approach that generates a map for each class to describe which features were most activated for predicting the class itself. In general, CAM based WSSS use the maps as ``seeds'' for predicting a first ROI on the object to be segmented. Then the seed ROI is used as a weak mask to train the proper semantic segmentation model~\cite{lin_broadcam_2023}. 

ROI identification through weakly supervised techniques is the best option for the diagnostic purpose because the physician is not hurried in precise pixel level analysis; rather she or he is simply alerted that some interesting thing may be present in a certain zone of the image. Moreover, the model itself does not need to be extremely precise, and this is an advantage in the medical imaging scenario where few training data are at disposal. In view of the previous considerations we decided to design our AI-supported diagnostic scenario as a ROI proposal one. The proposed workflow is agnostic with respect to the use of either bounding boxes or generic ROI contour because, in any case, the vertexes of the ROI can be properly stored in a DICOM RT Structure Set. Moreover, there is no difference in using IODeep to describe a network that learns bounding boxes or a model trained to predict ROIs. Finally, our workflow prescribes that each ROI/bounding box has to be validated manually by the physician; thus we ensure clinical explainability of the whole process.

\section{Motivation of the work}\label{motiv}

Even if it is well known from the literature that radiologists would like\break(semi-)automated decision support systems provided that they are non-invasive \cite{jorritsma2015}, DL models have not entered the routine diagnostic workflow due to lack of standardization. Indeed, the clinical community started addressing this issue from the DICOM data harmonization viewpoint with the aim of introducing a proper use of DICOM metadata for mixing established data sets and true clinical images to enable development, validation, and clinical translation of AI tools in a standardized way \cite{fedorov2023national}.
We claim that the main way to standardization is including DL in the DICOM standard. Such an integration would allow for the development of consistent and interoperable diagnostic tools across heterogeneous imaging platforms, making it easier for healthcare providers to integrate AI into their workflow. Second, a standardized framework for training and validating DL models is the only way to design distributed computing infrastructures where federated learning can be enabled. As well debated in the literature \cite{darzidehkalani2022federated,guan2023federated} federated learning is the key for implementing very effective models that are trained on adequate amounts of data thus providing high accuracy for a plethora of diseases, while scientific data sets are almost limited as regards both their size and the clinical goal they are designed for. In turn, very accurate models can be used effectively by XAI techniques to assess the most relevant image features which guided the model to a certain prediction \cite{van2022explainable}.

Recent literature reports several frameworks in the field of medical imaging that can be integrated with the DICOM standard. The work in \cite{Jesus_Bastiao_2023} introduces a vendor-independent platform that offers interfaces aimed at managing digitized slides and medical reports. It provides digital image analysis services compatible with other current standards. The solution incorporates the open-source Dicoogle PACS plugin architecture for interoperability and extensibility, enabling  customization of the proposed solution. In \cite{Lajara_2019} the Visilab Viewer is presented. It is a web-based platform compliant with the DICOM standard web architecture to solve the Whole Slide Image (WSI) format visualization from multi-frame DICOM images. 

The literature reports also several frameworks that make use of the DICOM standard to achieve a sort of ``loose'' AI integration into the clinical workflow rather than trying to include them in the standard itself.  In recent years, a great effort has been done for DICOM standardization of digital pathology where particularly effective pan-and-zoom characteristics are needed to visualize WSI data in a PACS system along with the connection to external AI applications.
A fully DICOM implementation of a platform for digital pathology has been proposed in \cite{HERRMANN201837} where both visualization and ML applications can connect using the DICOM PS3.18 RESTful web services (DICOMweb\texttrademark)\footnote{\url{https://dicom.nema.org/medical/dicom/current/output/html/part18.html}}. In this architecture external applications have to implement the DICOMweb client API that connects to the DICOMweb server using the HTTPS protocol. Moreover, the authors selected a suitable set of DICOM metadata to accomplish the clinical requirements needed for diagnosing microscope images.
In~\cite{10.1016/j.future.2022.10.025} the open decentralized platform devised by the EMPAIA Consortium is illustrated that enables pathologists to connect transparently using a web browser both to a medical and a computing infrastructure to manage their data while running third-party AI applications on the same data. The platform provides an abstraction on the DICOM infrastructure, and it is intended to be agnostic with respect to both PACS and WSI scanner vendors.

In \cite{Dikici2020} a roadmap is proposed for integrating AI image analysis algorithms into existing radiology workflows: a case study on brain MRI is presented. In \cite{Kathiravelu2021} a framework called \textit{Niffler} is presented, that is a Machine Learning (ML) framework retrieving images from the PACS using DICOM network listeners. Niffler extracts and processes metadata from the acquired images at the research clusters, then it executes both ML and real-time analytics pipelines on the radiological images, and their textual metadata. In \cite{Castillo2020} a ML model for the automatic analysis of medical images in the radiological workflow is integrated into a PACS system by using DICOM services provided by open-source tools. A DL architecture is trained for classifying chest X-ray images and it is reported as a case study. A DICOM Imaging Router is presented in \cite{dicomrouterbodypart2021} that incorporates CNNs for categorizing unknown DICOM X-ray images into five anatomical groups: ``abdominal'', ``adult chest'', ``pediatric chest'', ``spine'', and ``others''. 

The frameworks outlined above, connect themselves to a DICOM infrastructure to carry out their task. Therefore, their use requires many tricks that radiologists must learn \cite{Recht2020}. In essence, the DICOM standard and the AI techniques remain two separate worlds that play their role in synergy when needed, without real integration. 

In \cite{Rufenacht__2023} the PyRaDiSe package has been developed, which goes in the direction of a tight integration. PyRaDiSe is an open-source Python package which is independent of DL frameworks, and addresses the issue of artifacts caused by 2D reconstruction as it provides a framework for developing auto-segmentation solutions that can directly operate on DICOM data. Authors claim that  PyRaDiSe helps to bridge the gap between data science and clinical radiotherapy by facilitating the implementation of deep learning segmentation models in clinical research practice. Actually, PyRaDiSe has the same research objective as IODeep, and the authors leverage DICOM RT Structure Sets (RTSS) to allow data conversion from DICOM to other image formats thus enabling easy auto-segmentation routines. It is well known in the DICOM related literature~\cite{clunie2000dicom} that DICOM RTSS are a suitable place to store AI related information like the labels for models training, and also the framework we devised for IODeep makes use of DICOM RTSS to store ROI contours. Differently from PyRaDiSe, in IODeep there is no need to code the segmentation solution. IODeep provides the physician with predictions about relevant ROIs in a completely transparent way.

Right now, the information architecture of the DICOM standard does not foresee any type of inclusion of AI, neither as a proper IOD nor in terms of defining a suitable Information Object Module (IOM) to be included in a more general structure. Currently, DICOM is moving towards the integration of AI applications, and a Work Group has been created purposely that is the ``WG-23: Artificial Intelligence/Application Hosting''\footnote{\url{https://www.dicomstandard.org/activity/wgs/wg-23}}. The main activity of the WG-23 has been oriented in defining mechanisms for discovering heterogeneous AI services that can expose a suitable manifest for declaring the DICOM services provided to the imaging platform. In the present work we adopt an approach that relies on a direct extension of the DICOM information architecture. This architectural choice derives from the previous experience by some of the authors in developing a framework for adaptive configuration of the PACS viewers' GUIs based either on the content (i.e. reason for study, modality and body part) of the images to be displayed or on explicit preferences issued by the radiologist. Configuration information is stored in a dedicated IOM which extends the DICOMDIR IOD \cite{gambino2018}.
We think that the design of new components of the DICOM information architecture, as in the case of IODeep, makes the extension to the standard simpler than defining the interfaces to interact with an external software ecosystem. Moreover, the architectural solution devised by the WG-23 could be prone to problems in a training and/or fine-tuning scenario, when a huge amount of data would have to be moved across the interface to feed the DL model. 
We developed both a ``monolithic'' PACS client that implements all the workflow related to the use of IODeep and the service architecture that makes use of IODeep according to the indications of the WG-23. In the next section we detail our implementation, and we compare the two solutions.

\section{IODeep architecture}\label{arch}
\begin{figure}[!h]
    \centering
    \includegraphics[width=\textwidth]{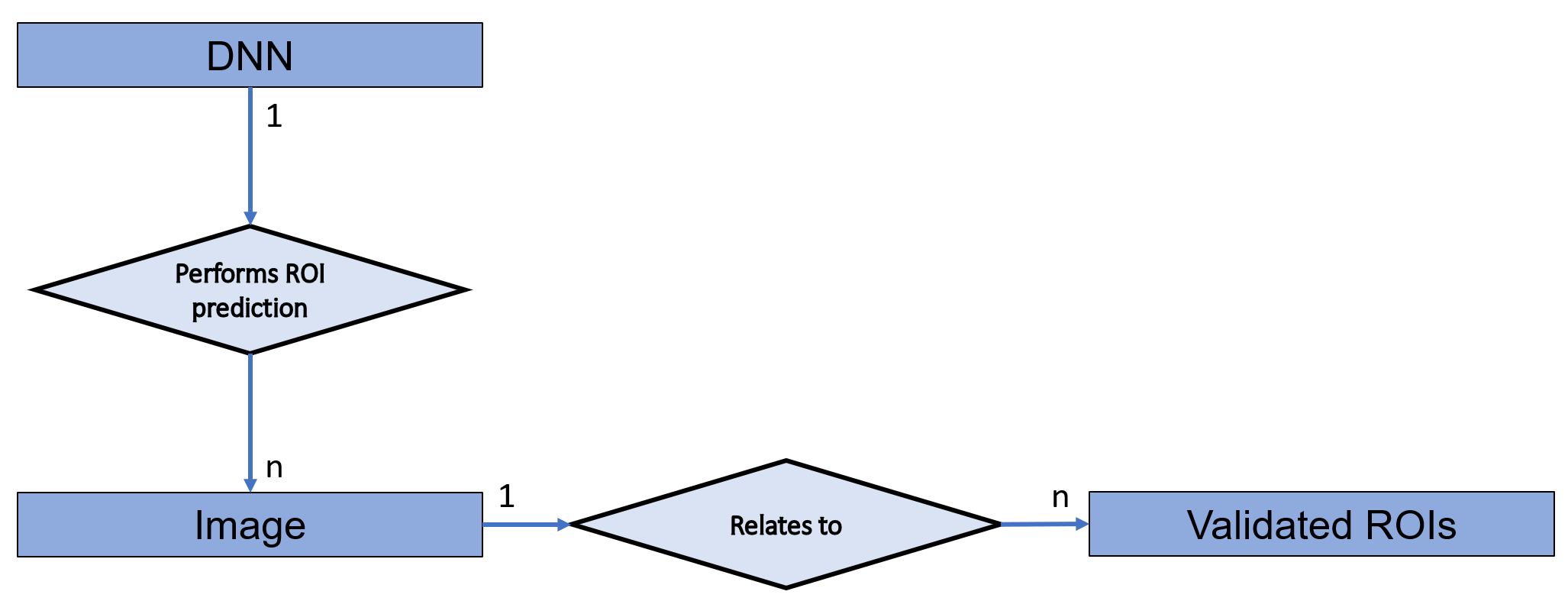}
    \caption{E-R diagram representing the connection between IODeep and the Dicom Model of the Real World}
    \label{figure_e_f_diagram}
\end{figure}

In figure  \ref{figure_e_f_diagram}  we report the E-R diagram describing how IODeep is connected to the DICOM Model of the Real World\footnote{\url{https://dicom.nema.org/medical/dicom/current/output/html/part03.html\#chapter\_7}}. As it can be noticed, the information contained in IODeep is used to instantiate a suitable DNN to predict ROIs on different images. In turn, references to such images are used to generate and store ROI data in the PACS infrastructure.

The overall IODeep structure is reported in table \ref{tab_tag_iodeep}.  The operations related to the selection of the proper neural model for the image under investigation, and making the actual prediction, rely on the information contained in four IOMs: ``DNN'', ``Image Pixel'', ``General Study'', and ``General Series''. The ``DNN'' module is a brand new one, which has been defined purposely to store the core info about the architecture, and the weights of the network.
Both the ``Image Pixel'' and the ``General Series'' module provide information to the network selection algorithm, but some tags in the ``Image Pixel'' module contain also information about image data formatting to give the DNN input tensor the correct shape. Finally, the ``General Study'' module is mandatory because the \textit{StudyInstanceUID} has to be filled with a proper value to create the Service Object Pair (SOP) Class aimed at storing/retrieving IODeep instances from the PACS server.

\begin{table}[!h]
    \centering
    \caption{The general \textit{IODeep} structure reporting both the modules and the tags required for selecting and instancing a DNN architecture}
    \resizebox{\columnwidth}{!}{\begin{tabular}{l|c|c|c}
    \hline
    Name&TAG& VR&Values\\
    \hline
    \multicolumn{4}{c}{DNN module}\\
    \hline
    \textit{DnnArchitecture}&(0017, 00XX)&UT&DNN architecture\\
    \textit{DnnWeitghts}&(0017, 00X1)&UT&DNN weight\\
    \textit{DnnName}&(0017, 00X2)&PN&e.g. “Brain Tumor segmentaion Unet”\\
    \textit{DnnUID}&(0017, 00X3)&UI&UID*\\
    \hline
    \multicolumn{4}{c}{Image Pixel module}\\
    \hline
    \textit{PhotometricInterpretation}&(0028, 0004)&CS&\begin{tabular}{@{}c@{}}RGB, MONOCHROME1...\\(DICOM standard Section C.7.6.3.1.2)\end{tabular} \\
    \textit{SamplesPerPixel}&(0028, 0002)&US&3 or 1\\
    \textit{PatientOrientation}&(0020, 0020)&CS&\begin{tabular}{@{}c@{}} [“P”, “F”]; [“L”, “P”];[“L”, “F”] ... \\ (DICOM standard Section C.7.6.1.1.1)\end{tabular}\\
    \textit{PlanarConfiguration}&(0028, 0006)&US&0 or 1\\
    \hline
    \multicolumn{4}{c}{General Study module}\\
    \hline
    \textit{StudyInstanceUID}&(0020, 000D)&UI&UID*\\
    \hline
    \multicolumn{4}{c}{General Series module}\\
    \hline
    \textit{SeriesInstanceUID}&(0020, 000E)&UI&UID*\\
    \textit{Modality}&(0008, 0060)&CS&\begin{tabular}{@{}c@{}} CT, MR, PT ...\\ (DICOM standard Section C.7.3.1.1.1)\end{tabular}\\
    \textit{BodyPartExamined}&(0018, 0015)&CS&\begin{tabular}{@{}c@{}} BREAST, ABDOMEN, CHEST ...\\ (DICOM Part 16: Content Mapping Resource)\end{tabular}\\
    \hline
    \end{tabular}}
    \label{tab_tag_iodeep}
    \footnotesize* These tags share the same Unique IDentifier that is the one defined for the \textit{DnnUID} tag
\end{table}

As required by the DICOM WG-23, IODeep is a non-Patient IOD, and it does not contain any patient information. Nevertheless, all the mandatory tags belonging to the ``Patient'' module (not reported in table \ref{tab_tag_iodeep}) along with the other modules composing IODeep, were included in its structure but they were left empty, as stated by the standard. The \textit{StudyInstanceUID} in the ``General Study'' module and the \textit{SeriesInstanceUID} in the ``General Series'' module are a mandatory tag, but they are not useful in the proposed architecture, and we decided to set them to the same value as the \textit{DnnUID} tag.

The ``DNN'' module contains the information about the network. Apart from the obvious meaning of the \textit{DnnUID}, and \textit{DnnName} tags, both \textit{DnnArchitecture} and \textit{DnnWeights} contain textual information about the layers and the weights of the network respectively. No particular formatting has been indicated in our tag specification, and that's why we used the Unlimited Text (UT) Value Representation. One can imagine to fill these tags with either a structured text like JSON and XML or a simple URI pointing to a binary file containing network information according to either a Tensorflow or a PyThorch file format. The second option is viable because in our application scenario, the DNN back-end is provided through a computing infrastructure in the same network where the PACS is. One can imagine either a dedicated server or better an instance of the DNN back-end running on a medical workstation because it will be used only in inference mode thus the computational effort is very limited. In both cases, no security concerns arise because the DNN back-end does not need to connect to the Internet, and the network instanced through IODeep predicts the ROIs starting from an image that is already open on the PACS viewer. This ensures that no privacy issues arise because the radiologist is authorized to see patient's data, and the network uses only the values stored in the ``Frame Of Reference'' and ``Image Pixel'' modules of the slice.

\begin{figure}[!h]
    \centering
    \includegraphics[width=\textwidth]{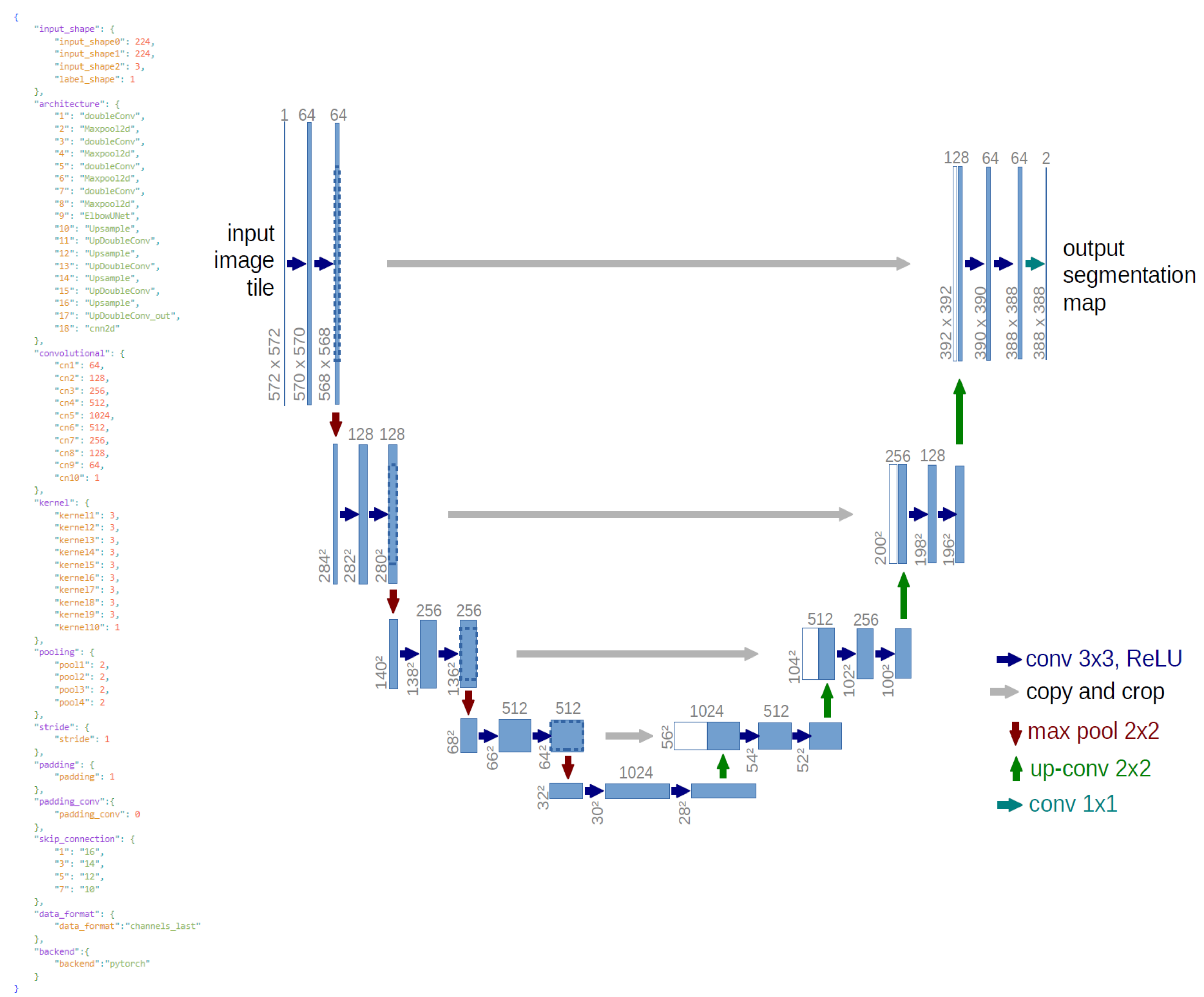}
    \caption{Representation of the U-net neural network used in our example, and its JSON description.}
    \label{fig_json_net}
\end{figure}

In our implementation the \textit{DnnArchitecture} tag contains a JSON document whose structure has been developed purposely by some of the authors along with the related parser, which allows to instantiate the same abstract model seamlessly in TensorFlow or PyTorch. The parser executable code is freely available in a suitable repository\footnote{\url{https://github.com/CHILab1/DNN-parser}} linked by the project's one. The repository contains also the command line specifications to invoke the parser, the complete documentation for writing user-defined JSON descriptions of a network architecture, and the JSON code for building the example U-net \cite{Ronneberger_2015} used in our implementation. The JSON code is showed also in figure \ref{fig_json_net} along with the U-net architecture. As it can be seen, the key \texttt{input\_shape} contains the information about the shape of the input tensor, which will be used to check if reshaping is needed for the slice to be processed. The key \texttt{architecture} shows the type and sequential order of the layers that make up the neural network, while the keys \texttt{convolution}, \texttt{kernel}, and so on report the parameters to be set to implement the convolution properly for each layer. In the example a simple sequential model is reported, but our parser deals with all the layer types both in TensorFlow and PyTorch. As an example, our format supports the key \texttt{skip\_connection} to build U-net models that are typical architectures for medical imaging. A detailed description of the implemented JSON keys, and the executable of the CLI version of the parser are available in the project repository.

In the proposed implementation, we used a PyTorch back-end so the tag \textit{DnnWeights} contains the URI pointing to a file with \texttt{.pth} extension that is the binary format used by Torch, storing the weights as a serialized Python object. In Tensorfllow we adopted the \texttt{tf} save format to allow the implementation of custom layers. We choose to save separately the weights and the architecture to allow simple creation of the network by the user, who does not need to have strong coding skills. When running the parser, either a \texttt{torch} or a \texttt{tensorflow} environment is set according to the value passed in the \texttt{--backend} command line parameter. The environment is used just to instantiate a specific model that will be saved using the path specified in the \texttt{--name} command line parameter. In this way, a network developer has no need to instantiate a specific model class inherited from either \texttt{torch.nn.Module} or \texttt{tensorflow.keras.Model}. The dcm file containing the IODeep used to instantiate our U-net example network is available in the project repository.

\subsection{The ROI prediction workflow}
\begin{figure}[!b]
    \centering
    \includegraphics[width=0.73\textwidth]{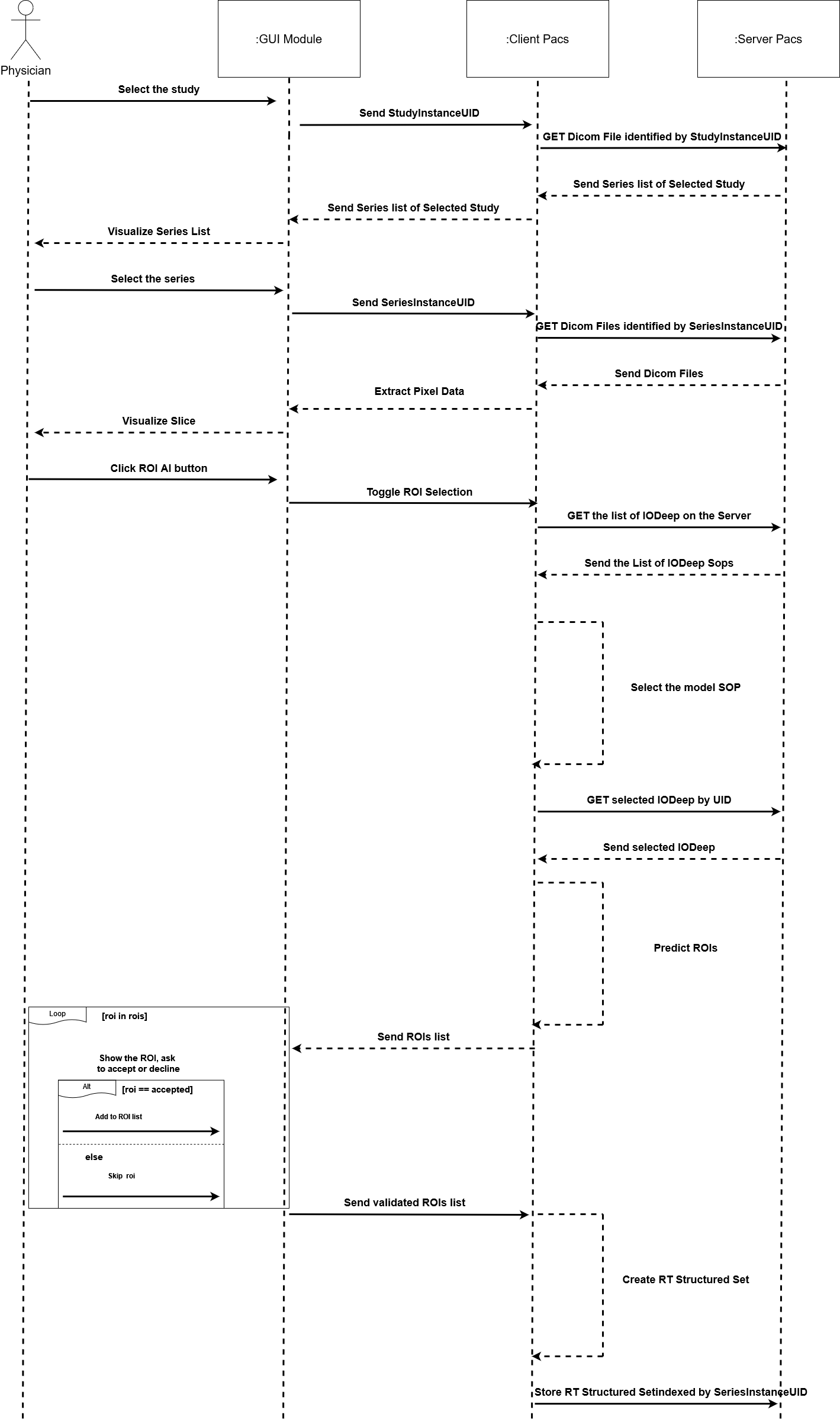}
    \caption{Sequence diagram of the ROI prediction scenario}
    \label{fig_sequence_diagram}
\end{figure}

Figure \ref{fig_sequence_diagram} reports a simple UML sequence diagram showing the workflow of the ROI prediction scenario. We developed a lightweight PACS client to show the effectiveness of this workflow.

All the IODeep instances corresponding to networks that can be served by the DNN back-end are within the PACS server, and they can be made available to the physician who wants to perform automatic prediction of the ROIs in a particular slice. To test our implementation, we trained a couple of example Unets on the ``Brain Tumor Classification (MRI)'' data set\footnote{\url{https://www.kaggle.com/datasets/sartajbhuvaji/brain-tumor-classification-mri}} and the ``UW-Madison GI Tract Image Segmentation'' data set \footnote{\url{https://www.kaggle.com/competitions/uw-madison-gi-tract-image-segmentation/overview}}respectively. 

After the study request, and the visualization of the series to be diagnosed, the physician can click on a suitable ``AI ROI" button at the viewer GUI while she or he is viewing whatever slice in the series. This event will start the actual ROI prediction process. We implemented a thick PACS client that performs both the network selection and the true ROI prediction by instancing the network on the DNN back-end. In this scenario, the client reads the tags stored in the ``General Series'' and ``Image Pixel'' modules inside the Image IOD of the current slice, and it retrieves all the IODeep instances from the server to start network selection. 

Specifically, the network selection algorithm will use the \textit{Modality} tag (0008, 0060) which describes the type of data acquisition, the \textit{BodyPartExamined} tag (0018, 0015) which describes the anatomic region of the study using Code Strings (CS) that refer to the standardization reported in the DICOM Part 16: Content Mapping Resource, and the \textit{SamplePerPixel} tag (0028, 0002) which describes the number of channels in the data. In case the \textit{BodyPartExamined} tag is empty, the content  of the \textit{StudyDescription} tag (0008, 1030) is used. The matching algorithm for selecting the neural network is shown below in Algorithm \ref{alg_match}.

\begin{algorithm}[!ht]
\caption{Network selection algorithm}\label{alg_match_nn}
\scriptsize\begin{algorithmic}[1]
\Require $SliceTagList$:\,list($Modality$,\,$SamplePerPixel$, \,$BodyPartExamined$,\,$StudyDescription$) \Comment{The list of relevant tags from}
\Statex\hspace{71mm}the current slice
\Require $IODeepList$ \Comment{The list of all the IODeep instances retrieved from}
\Statex\hspace{38mm}the PACS server
\If{$SliceTagList.BodyPartExamined == \emptyset$}
    \State $SliceTagList$.remove$(BodyPartExamined)$
\Else{}
    \State $SliceTagList$.remove$(StudyDescription)$
\EndIf
\For{$net$ in $IODeepList$}
\State $IODeepTagList \gets$ \,list($net.Modality$,\,$net.SamplePerPixel$,
\Statex\hspace{43mm}$net.BodyPartExamined$)
\State $control \gets 0$
\For{$slice\_tag, net\_tag$ in ($SliceTagList, IODeepTagList$)}
\If{$slice\_tag$ == $net\_tag$\,and\,not\,$SliceTagList$.isLast($slice\_tag$)}
\State $control \gets control + 1$
\EndIf
\If{$SliceTagList$.isLast($slice\_tag$)\,and\,($slice\_tag$\,==\,$net\_tag$\,or
\Statex\hspace{15mm}$slice\_tag$.substring($net\_tag$)\,==\,True)}
\State $control \gets control + 1$\Comment{When using \emph{StudyDescription} a pattern} \Statex\hspace{55mm}search is used in place of strict equality
\Statex\hspace{55mm}on the last tag
\EndIf
\EndFor
\If{control == 3}
    \State $net\_sop\_uid \gets net.DnnUID$\Comment{the SOP UID of the selected IODeep} \Statex\hspace{58mm}instance
    \State\Return $net\_sop\_uid$
\Else{}
    \State $net\_sop\_uid \gets None$
\EndIf
\EndFor
\State\Return $net\_sop\_uid$
\end{algorithmic}
\label{alg_match}
\end{algorithm}

In case of successful match, the SOP UID of the IODeep is returned by the algorithm. If no IODeep instance satisfies requirements for network prediction, a warning message will be returned blocking the work pipeline. The returned SOP UID will be used to query the PACS server for obtaining both the network architecture and the URI pointing to the weights file. In turn, the JSON document in the \textit{DnnArchitecture} tag will be parsed to create and save the network instance using the DNN back-end, which is selected according to the value of the \texttt{backend} key. The parser itself is invoked by the PACS client using a system call. After the the input tensor shape check, actual prediction will be made. The entire ROI prediction workflow is reported in the Algorithm \ref{alg_roi_pred}.

Before the actual creation of the network instance in the DNN back-end, the shape of the current slice is checked to assess if it is compatible with the network input tensor. The tags \textit{SamplePerPixel} (0028, 0002), \textit{Rows} (0028, 0010), and \textit{Columns} (0028, 0011) are used to check if the dimensions of the image are compatible with the neural network input or if reshape is required. Moreover, The tag \textit{PhotometricInterpretation} (0028, 0004), is used to verify the interpretation of the given pixels (e.g. MONOCHROME2, MONOCHROME1, RGB, and so on). Finally, the tag \textit{PixelRepresentation} (0028, 0103) describes the pixel representation (signed or unsigned) and will be the last tag checked for the final formatting of the image to be tested. This process is summarized by the \texttt{check\_tensor\_shape()} call in the trace of Algorithm \ref{alg_roi_pred}.

Once the shape of the image has been checked w.r.t. the expected network input tensor, the actual model is created in the DNN back-end starting from the object parsed from the JSON document in the \textit{DnnArchitecture} tag, and the corresponding weights are loaded. 
The actual ROI prediction is a very fast process, which will output a list containing all the poly-lines that have been predicted.

\begin{algorithm}[!h]
    \caption{ROI prediction algorithm}\label{alg_roi_pred}
    \scriptsize\begin{algorithmic}[1]
    \Require $IPModule$ \Comment{The ImagePixel module of the current slice}
    \Require $FRModule$\Comment{The FrameOfReference module of the current slice}
    \Require $IODeep$ \Comment{The selected DNN}
    \Require $Backend$\Comment{The reference to the DNN back-end}
    \State $rois \gets$ list()
    \State $network \gets$ parser($IODeep.DnnArchitecture$)
    \State check\_tensor\_shape($IPModule$,$network.input\_shape$)\Comment{tensor shape}
    \Statex\hspace{97mm}analysis
    \State $model \gets$\,$Backend$.createNetwork($network$)\Comment{instance of the network in}
    \Statex\hspace{75mm}the DNN back-end
    \State $model$.load($IODeep.DnnWeights$)\Comment{loading method of the DNN back-end}
    \State $preds \gets$ $model$.predict($IPModule.pixelData$)\Comment{prediction method of the}
    \Statex\hspace{77mm}DNN back-end
    \For{$roi$ in $preds$}:
        \State $slice\_id \gets FRModule.FrameOfReferenceUID$
        \State $rois$.add(dict($``sliceID":\,slice\_id,\,``polyline":\,roi$)\Comment{These value will} 
        \Statex\hspace{36mm}populate the RT Structure Set containing the ROIs
    \EndFor
    \State \Return $rois$
    \end{algorithmic}
\end{algorithm}

\begin{figure}[!h]
     \centering
     \begin{subfigure}[b]{0.47\textwidth}
         \centering
         \includegraphics[width=\textwidth]{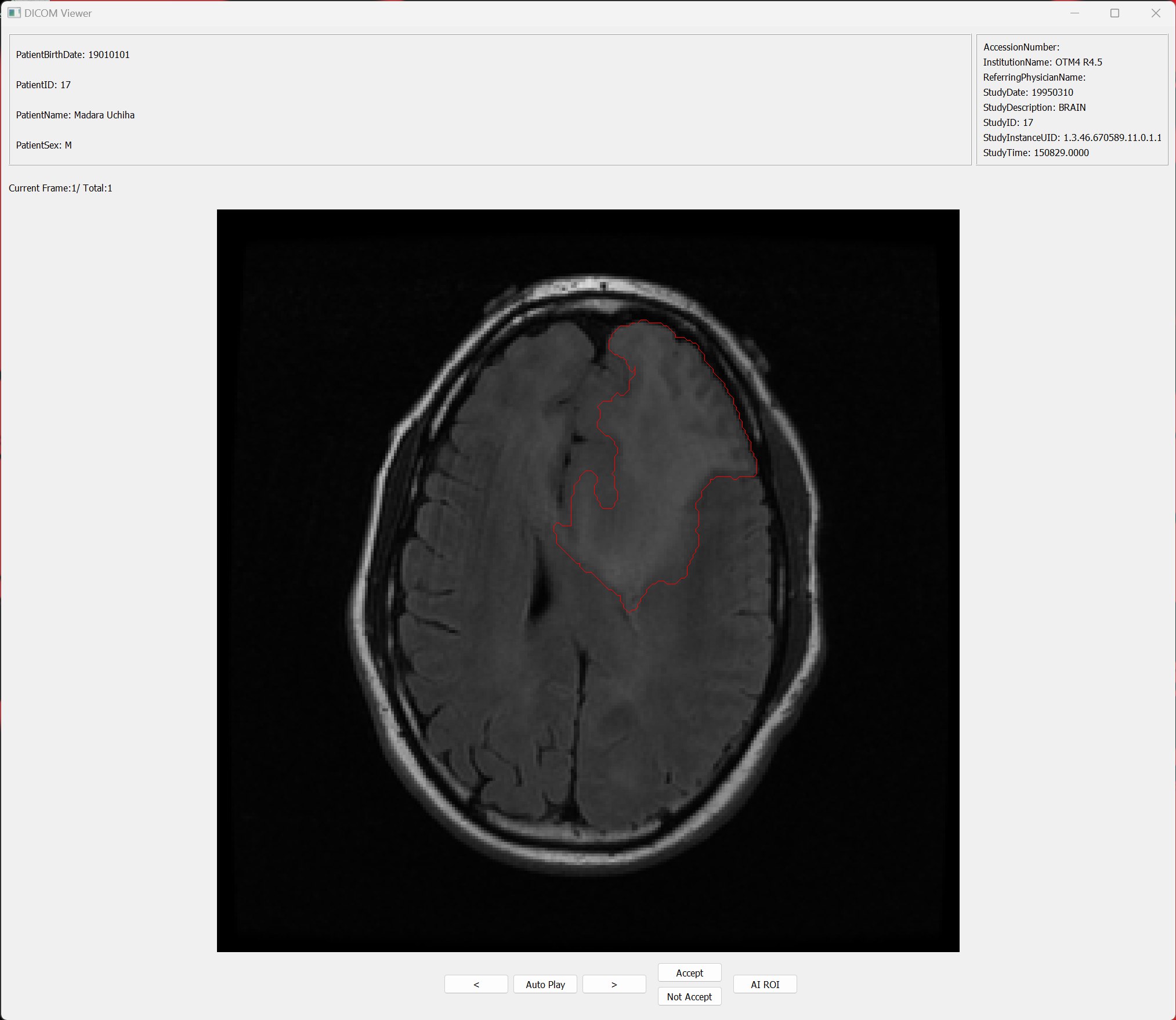}
         \caption{}
         \label{fig_predicted_roi}
     \end{subfigure}
     \begin{subfigure}[b]{0.47\textwidth}
         \centering
         \includegraphics[width=\textwidth]{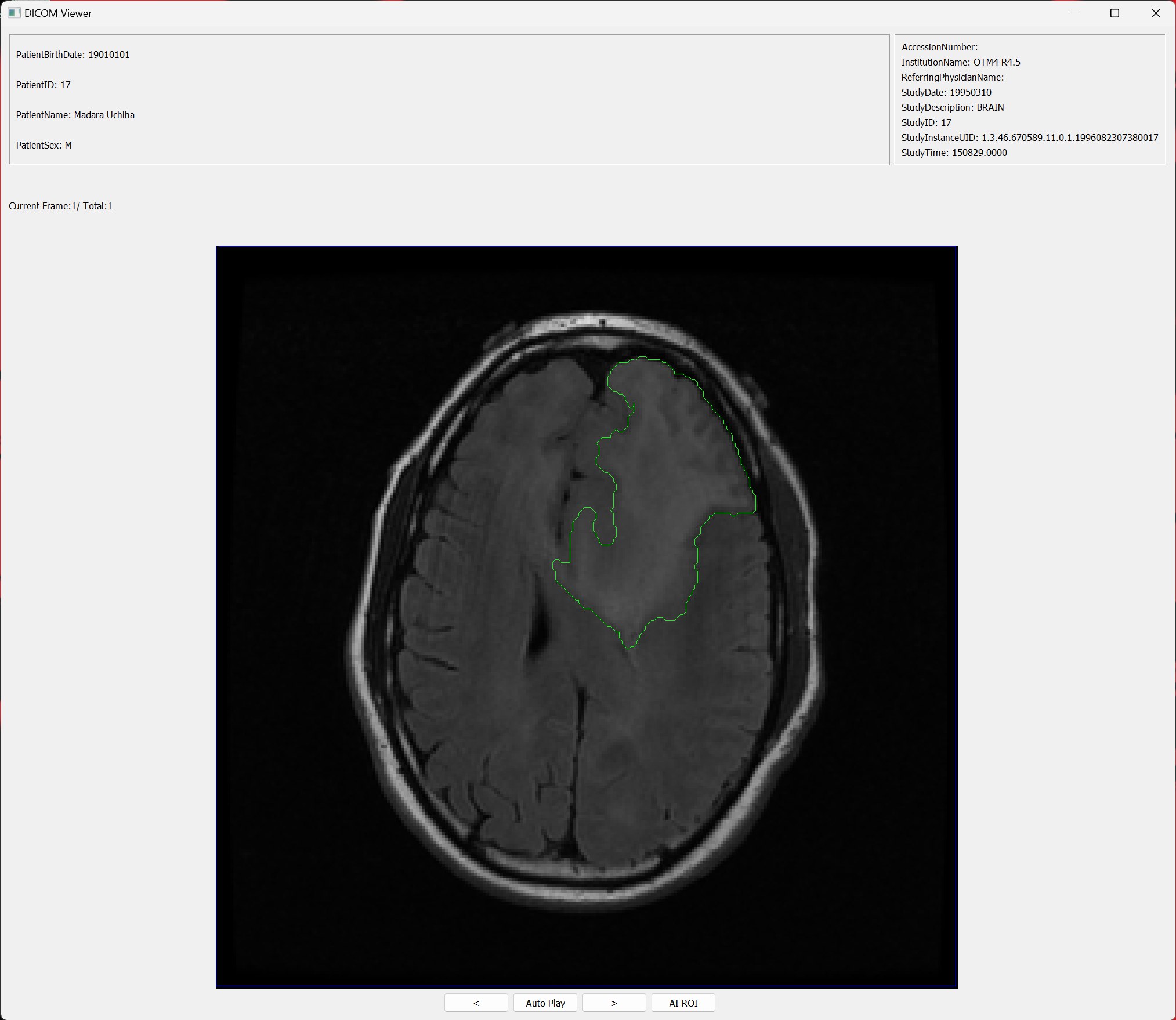}
         \caption{}
         \label{fig_validated_roi}
     \end{subfigure}
        \caption{The viewer interface for ROI validation. (a) Predicted ROIs are displayed and outlined in red. (b) Validated ROIs are outlined in green.}
        \label{fig_viewer}
\end{figure}

Each ROI will be displayed by the viewer, giving the physician the option of either validating or rejecting the region. Figure \ref{fig_viewer} shows how the ROI can be validated in our simple viewer that was implemented using the Python library PyQt. The viewer interfaces with the PACS server, and returns to the user a list of the available studies. Once a study has been selected, a list of that study's series is retrieved in order to select the slice to be viewed. The display will show two information blocks in the upper section. The left block displays patient's information: \textit{PatientName} (0010, 0010), \textit{PatientID}(0010, 0020), \textit{PatientBirthdate} (0010, 0030) , \textit{PatienSex} (0010, 0040). The second block displays clinical information: \textit{AccessionNumber}(0008, 0050), \textit{InstitutionName}(0008, 0080), \textit{ReferringPhysicianName} (0008, 0090), \textit{StudyDate} (0008, 0020), \textit{StudyDescription} (0008, 1030), \textit{StudyID} (0020, 0010), \textit{StudyInstanceUID} (0020, 000D), and \textit{StudyTime} (0008, 0030). 
The central body of the viewer will display the image. In this part of the layout there are the ``autoplay'' and the navigation buttons apart from the ``AI ROI'' button, already mentioned. The interaction between the viewer and the DICOM standard was developed using the Pydicom library, while the server was an instance of the Orthanc DICOM server\footnote{\url{https://www.orthanc-server.com/}}.

The ROIs validated by the physician will be stored within a DICOM RT Structure Set, a data structure used to store all information that is collected through virtual instruments/workstations. In this data structure, all the patient information contained in the source IOD will be kept. Moreover, both the ``Frame of Reference'' and the ``ROI Contour'' module will be added to instance the \textit{FrameReferenceUID} (0020, 0052) and the \textit{ROIContourSequence} (3006, 0039) tag respectively, which contain the slice reference and the actual ROI data. Finally, the validation information will be stored through the ``Approval'' module that will contain the tags:

\begin{itemize}
    \item \textit{ApprovalStatus} (300E, 0002)
    \item \textit{ReviewDate} (300E, 0004)
    \item \textit{ReviewTime} (300E, 0005)
    \item \textit{ReviewerName} (300E, 0008)
\end{itemize}

The RT Structure Set will be uploaded to the PACS server to be reused should the physician wish to retrieve ROI information. The DICOM file containing the RT Structure Set generated from the ROI displayed in figure~\ref{fig_viewer} is available in the project repository. As already pointed out in section \ref{sec:back}, even if we used an example U-net trained at predicting ROIs, IODeep can be used equally to instantiate both networks that deal with ROIs and models trained at devising bounding boxes. In any case the DICOM RT Structure Set will store the vertex list used to render  the poly-line contour in the PACS viewer.

\subsection{Comparison with the DICOM WG-23 proposal}
People in the DICOM community are working actively towards the integration of AI algorithms in the standard, and a suitable Working Group has been founded in this respect: the WG-23 ``Artificial Intelligence and Application Hosting''. The current proposal of the WG-23 has been submitted in the form of \emph{work item}\footnote{\url{https://www.dicomstandard.org/workitems}} that is the standard way for working groups to publish their results. 

The core of the WG-23 work item is to pursue AI integration using a mechanism of (micro-)service discovery according to the model proposed by the Open Application Model (OAM)\footnote{\url{https://oam.dev/}}. In this model, AI applications are deployed as services running on heterogeneous computing infrastructures that can be either on the same network of the PACS or in an external cloud. Each application owns a ``Manifest'' in the form of a YAML file, which encapsulates the definition of a workload and the information needed to run it. OAM defines also an ``application operator'' that is the infrastructure that actually deploys the application, and defines more information from the operational point of view to allow the so called ``Platform'' (the PACS infrastructure using the services, in our case) to integrate them explictly in its workflow. In the vision of the WG-23, applications can either register to the Platform or a they can be detected tyhrough a service discovery mechanism.

Our proposal moves from a monolithic implementation perspective where information is fully specified according to the DICOM Information Architecture, and only a TCP/IP socket connection is needed to reach the DNN back-end that runs on the same network of the PACS. Apart from the privacy and security concerns already mentioned, we adopted this solution for efficiency reasons. The services ecosystem proposed by the WG-23 is a scalable solution with very low implementation effort on the PACS side, but it is tailored on an inference scenario, and it is not well suited to perform training or fine-tuning. In general, Radiology wards are focused to particular diseases due to their location in the territory, and the presence of precise medical specializations. As a consequence, fine tuning a trained network by means of the peculiar data generated in a particular PACS infrastructure, appears a reasonable extension of using DNNs in Medical Imaging. Making a data set available at the DNN back-end for training a model is a bandwidth consuming task. A service implementation is not well suited for training and/or fine tuning because, in general, a service infrastructure is intended to be hosted in a remote cloud with respect to the applications that require a service. In the model training scenario, this leads to both high latency and privacy issues. On the other hand, our solution requires only a data transfer on the internal network of the ward, and this is not so resource consuming. IODeep can be used explicitly also in this context: a PACS client devoted to back-end administration could select a data set made by several slice series, and start a network selection procedure, that is the same we described before, followed by the training in the back-end.  

\begin{figure}
    \centering
    \includegraphics[width=0.8\linewidth]{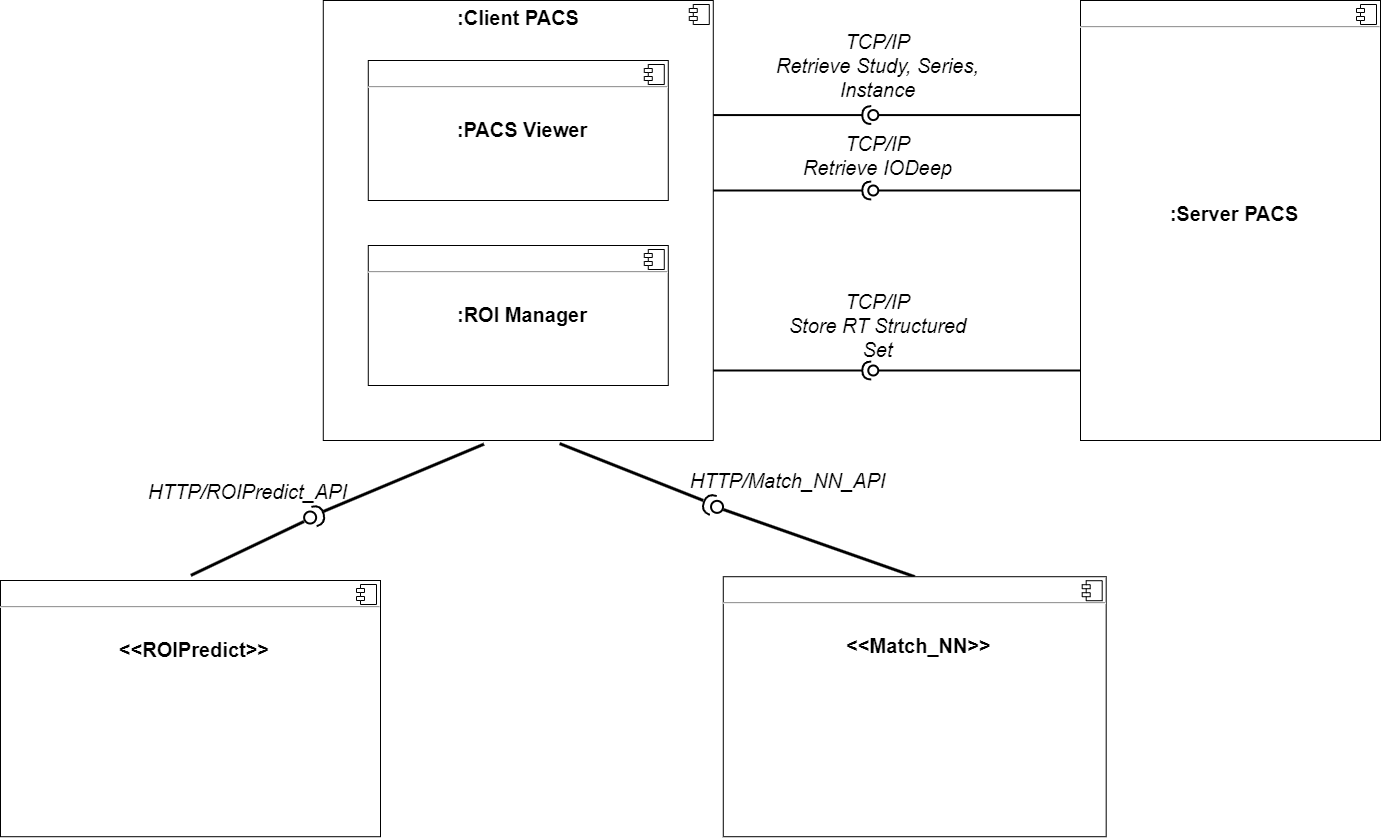}
    \caption{Component diagram representing the service implementation of the proposed framework.}
    \label{fig:service-architecture}
\end{figure}

Nevertheless, we rearranged our implementation according to the indications of the WG-23 work item. The software architecture of this solution is reported in figure~\ref{fig:service-architecture}. In this scenario, Algorithm~\ref{alg_match_nn} has been redesigned to provide a service discovery REST API (Application Programming Interface) and it is available at a known endpoint on the service infrastructure. Such an URI is the only thing the PACS platform has to know, and IODeep stores the information used to call the API at the service discovery endpoint. This interaction provides the platform with the endpoint of the selected DNN service. The PACS platform calls the DNN service API at the service endpoint passing the slice, and the information for checking the tensor shape stored in the selected IODeep. Finally, the selected DNN service runs the Algorithm~\ref{alg_roi_pred}. This service architecture does not need to store network information in the IODeep architecture as we can assume that all the networks are already instanced in the DNN back-end. In our implementation the PACS server, the client, and each REST service run in separate Docker containers that communicate throughusing a virtual LAN defined through the YAML configuration file. The actual containers can be deployed on either a single machine or a cluster. We choose explicitly this kind of infrastructure because we strongly believe that effective extension of the service architecture to the training/fine-tuning scenario can be achieved only in a LAN context where the concerns related to the connection bandwidth are negligible.

\section{Conclusion}\label{conc}

In this paper, we introduced IODeep, a new DICOM Information Object Definition aimed at integrating all the information needed to allow a pre-trained DNN model to be selected w.r.t. the features of the slice under investigation, and to be instanced in a DNN back-end running in the same network of the PACS server to perform inference about ROIs in that slice.
We presented the detailed information architecture of IODeep along with a precise workflow for selecting the correct DNN, instancing the abstract model description in the back-end, making predictions, and storing the ROIs validated by the physician in a RT Structure Set for further reuse. Finally, we implemented a suitable PACS client to show the effectiveness of our proposal.

We strongly believe that the key for using Artificial Intelligence in the everyday diagnostic activity for medical images is related to standardization. Our approach provides a complete answer to the first statement in Section~\ref{sec:intro}, and it can be reviewed as a technological starting point for answering the other two statements. 

DNNs are useful if they can be fine-tuned on the real data sets available at the Radiology wards, and our solution allows a simple extension to deal with this scenario. In fact, the same information architecture in IODeep can be used to select a model to be retrained provided that the bandwidth concerns to move data from the PACS to the DNN back-end are easy to solve. In this respect, a solution that runs entirely in the same LAN is strongly preferable to a services ecosystem running in cloud.

We also showed that IODeep can be used effectively to implement a service based solution for AI integration in DICOM, as the one devised by the WG-23. In fact, our network selection and ROI prediction algorithms can be implemented as a service discovery procedure, and DNN service respectively to pursue this goal.

Many DNNs for Medical Imaging work on multi-modal data: this is the case of registered CT-PET volumes, or the very recent generative AIs that integrate both textual and visual information. The current implementation of IODeep does not support multi-modal input data, but this is not a strong limitation for two reasons. First, the general two-stage approach based on both DNN selection and ROI prediction does not depend on the particular input data. Second, the structure of IODeep in Table~\ref{tab_tag_iodeep} can be easily extended to utilise multi-modal and/or textual data. In this respect, one can devise a ``Study IODeep'' that is the proper IOD containing the tags related to both the architecture and the weights of the DNN. Moreover, different ``Series IODeep'' modules will contain the information related to each kind of input tensor, which can be either a visual or a textual one. Each ``Series IODeep'' module will reference the \textit{DnnUID} of the ``Study IODeep'' to allow proper retrieval of all the information needed to select the multi-modal network, and to check the shape of all the input tensors. We are now working actively on this issue.

Finally, we want to point out that IODeep is also a technological key for explainability of DNN models, and their acceptance in the radiological community. First of all, we claim the need to support the diagnostic process just with a ``ROI suggestion'' task.  A model suggesting a loose ROI does not make a precise pixel level segmentation. In fact, validating an exact segmentation represents a huge cognitive load for the doctor. Moreover, precise ROI prediction can be achieved only with extensive training on very huge amounts of data. Such a goal can be pursued using Federated Learning (FL) that is a distributed learning paradigm where the model is trained while ``moving'' it across the data repositories where the data set is split, but data are always kept where they reside without moving them at all. The technology proposed in this paper enables a distributed scenario where IODeeps are shared between the PACS servers of various public and private hospital institutions without violating any rule related to either data security or patients' privacy, while allowing FL through the data stored within the individual PACS servers.

\section*{Author contributions}

SC and LC contributed equally to research activity. They performed Data curation, Investigation, Methodology, Software and Writing-original draft. OG performed Conceptualization and Writing-original draft. RP supervised the research team made Writing-review $\&$ editing, and contributed to Conceptualization, Formal analysis, Project administration. 

\bibliographystyle{plain}
\bibliography{mybibfile}

\end{document}